# Combination of Optical Transitions of Polarons with Rashba Effect in Methylammonium Lead Tri-halide Perovskites under High Magnetic Fields

Y. H. Shin,[1] Halim Choi,[1] C. Park,[1] D. Park,[2] M. S. Jeong,[2] H. Nojiri,[3] Z. Yang,[4] Y. Kohama,[4] and Yongmin Kim[1]*

[1] *Department of Physics, Dankook University, Cheonan 31116, Korea*
[2] *Department of Physics, Hanyang university, Seoul 04763, Korea*
[3] *Institute for Materials Research, Tohoku University, Sendai 980-8577, Japan and*
[4] *Institute for Solid State Physics, The University of Tokyo, Kashiwa 277-8581, Japan*

(Dated: April 26, 2021)

We investigate photoluminescence (PL) transitions of $MAPbX_3$ (X = I, Br and Cl) organic-inorganic hybrid perovskite single crystals under magnetic fields of up to 60 T. In these materials, sharp free-exciton transition peaks emerge at a low temperature (4.2 K). Under strong magnetic fields, the free-exciton PL transitions of three different halogens show dramatic differences. The free-exciton transitions of the $MAPbCl_3$ crystal undergo negative energy shifts, while those of the $MAPbBr_3$ crystal show normal diamagnetic shifts. To obtain the variation from Cl to Br, we attempt to measure PL transitions of $MAPbCl_xBr_{3-x}$. For $MAPbI_3$, the transition-energy shifts for both $\sigma^+$ and $\sigma^-$ exhibit a power-law dependence on the magnetic field. Such inconsistent magnetic-field effects on different halogens make it difficult to understand the transition-energy behavior through a unified model. We propose a possible mechanism for the field effects that is based on a combination of the Rashba effect induced by strong spin-orbit coupling and the polaron effect caused by the polar nature of the inorganic elements.

## I. INTRODUCTION

Organic-inorganic hybrid perovskites (OIHPs) of the form $MAPbX_3$ (X = I, Br, and Cl) have been intensively investigated owing to their versatility for use in not only basic research but also industrial applications. Although vast investigations were initiated on $MAPbI_3$ because of its high solarcell efficiency of over 25%[1, 2], bandgap en- gineering by substituting or mixing with other halogens (Br and Cl) made it possible to extend its applications to light-emitting as well as light-detection devices. How- ever, even with the enormous potential for optoelectronic applications, the detailed physical characteristics of this material family remain unclear.

An indirect band alignment in the conduction and va- lence bands, which is necessary for a slow recombination rate, has been suggested, where the Rashba effect in- duced by the strong spin-orbit coupling (SOC) within its inorganic lattice framework plays an important role. The strong SOC of the Pb $6p$ orbital produces a large Rashba splitting in the conduction-band minimum at the $R$-point, whereas the weak SOC of the Pb $6s$ and halogen $p$ orbitals yields a small Rashba splitting in the valence band. The large and small Rashba splittings are pro- posed to be responsible for indirect band alignment in $MAPbX_3$. The Rashba effect in OIHPs is an intrigu- ing but challenging research subject. For example, the exciton fine structure has been investigated in $MAPbI_3$ nanocrystals. The excitonic state is known to split into ground singlet and excited triplet states owing to the ex- change interaction. The Rashba effect exchanges these states such that the triplet states become the ground state and the singlet states transform into a dark excited state[3–6]. Recently, it was suggested that dynamic and static Rashba effects could appear in bulk $MAPbBr_3$ and $CsPbBr_3$ crystals. The dynamic Rashba effect occurs in- side bulk crystals owing to the dynamic orientation of the $MA^+$ or $Cs^+$ cation in the inorganic cage. The effect dis- appears at low temperatures below 100 K for $CsPbBr_3$, whereas it persists below 100 K for $MAPbBr_3$. The static Rashba effect, on the other hand, emerges only on the surface of a bulk crystal and originates from the surface reconstruction through $MA^+$ cation ordering[7].

Due to the polar crystal characteristic of OIHPs, charge carriers interact with the phonons generated by the polar lattice, forming small and large polarons[8–11]. The polaron effects in OIHP are as complicated as their complicated structure. Both small and large polarons modify the effective mass and mobility of charge carriers because the carriers are localized near the polar lattice. It is generally known that the small and large polarons show an incoherent hopping and a band-like transport behavior, respectively. The dynamic orientation of the organic cation ($MA^+$) that couples with excess charges causes the distortion of the inorganic octahedron, form- ing a small polaron[10, 12, 13]. Unlike the small polaron, in which the charge carriers are localized within several unit cells, the charge carriers in the bands, coupled with the longitudinal optical (LO) phonon caused by the in- organic lattice bonding, generate a large polaron.[10, 14– 16]. The formation of the large polaron is due to the Pb- halide-Pb deformation modes and does not depend on the cation species (MA or Cs)[17]. Therefore, the substi- tution of halides is expected to be the origin of the differ-ence in the character of the large polarons. OIHPs with different halogens have different bond polarities because of the electronegativity difference. The known Pauling electronegativities of inorganic elements in halide perovskites are 1.87 (Pb), 3.16 (Cl), 2.96 (Br), and 2.66 (I). For a polar covalent bond, the fraction of ionic bonding can be calculated by using the following simple equation:

$$fraction\ of\ ionic\ bonding\ (\%) = (1 - e^{-(\Delta x/2)^2}) \times 100, \quad (1)$$

where $\Delta x$ is the electronegativity difference between the ionic elements, which are Pb and the halogen in OIHP. The calculated fractions of ionic bonding of the halide perovskites are 34.0% (MAPbCl$_3$), 25.7% (MAPbBr$_3$), and 14.4% (MAPbI$_3$). Such differences in the ionic bonding fraction of OIHPs may cause the polaron effects with the replacement of the halogen. Systematic Raman and infrared spectroscopy studies[18, 19] reported that with the substitution of the halogen in the sequence of Cl, Br, and I, the vibrational frequencies of the OIHP undergo red-shifts, which can be attributed to the reduction of the ionic bonding fraction. In Ref. [18], the authors found that irrespective of the MA cation species, element substitution in the sequence of Cl, Br, and I decreases all vibrational frequencies including stretching, bending, rocking, and torsion modes.

_________
*correspondingauthor:yongmin@dankook.ac.kr

In this work, we report magneto-photoluminescence (MPL) transitions of MAPbX$_3$ (X = Cl, Br and I) single crystals under high magnetic fields of up to 60 T by using capacitor bank driven pulsed magnets and up to 19 T by using a superconducting dc magnet at 4.2 K. Optical transition measurement under magnetic fields is a powerful tool to determine various physical parameters of given materials, such as the dielectric constant, effective mass, $g$-factor, exciton Bohr radius, and/or exciton binding energy. Most previous works involving magneto-absorption measurements under ultrahigh magnetic fields focused on the exciton Bohr radius and binding energy for OIHPs[20–22]. In the present study, we report that MPL transitions of free excitons (FX) strongly depend on the halogen in an inconsistent manner. For the MAPbCl$_3$ crystal, two FX transitions undergo spectral red-shifts, and a new peak emerges above 20 T, which shows a blue-shift as the magnetic field increases. For the MAPbBr$_3$ crystal, two FX transitions can be understood as ordinary diamagnetic energy shifts with different diamagnetic coefficient obtained from magneto-absorption measurements. To investigate spectral difference between MAPbCl$_3$ and MAPbBr$_3$, we measured MPL transitions of MAPbCl$_x$Br$_{3-x}$ ($x$ = 0.5, 1.0, 1.5, 2.0, and 2.5)crystals. For the MAPbI$_3$ crystal, the FX transition shows a power-law dependence on the magnetic field. Such unconventional optical transitions caused by the different halogens under magnetic fields are difficult to explain with one unified model and can be understood by the combination of the Rashba and polaron effects.

## II. EXPERIMENTAL SECTION

### A. Sample Preparation

All chemicals employed to grow samples were used as received without further purification. Methylammonium halides (CH$_3$NH$_3$X; X = Cl, Br, I) were sourced from Greatcell Solar Materials. Lead (II) halides (X = Cl, Br, I) were purchased from Alfa Aesar. Reagent-grade organic solvents such as $\gamma$-butyrolactone (GBL), dimethyl sulfoxide (DMSO), and dimethyl formamide (DMF) were purchased from Aldrich. MAPbX$_3$ (MA = CH$_3$NH$_3^+$ (methylammonium), X = Cl$^-$, Br$^-$, Cl$^-$) single crystals were prepared using the inverse temperature crystallization (ITC) method[23, 24]. For the precursor solution, the equivalent molar ratio of MAX and PbX$_2$ was dissolved in different solvents depending on the halide: DMSO/DMF (1:1, v/v) for Cl, DMF for Br, and GBL for I. The precursor solutions were kept with stirring for 1 day and filtered with a 0.2 $\mu$m PTFE syringe filter to obtain a clear solution. Organic lead halide perovskite single crystals were formed by keeping the filtered solution for 1 day at different temperatures: 60 °C for Cl, 80 °C for Br, and 110 °C for I. Crystals were washed with copious amounts of hexane to remove the residual solution on the surface and dried under vacuum for 1 day.

### B. Photoluminescence Setup

For temperature-dependent PL measurements, a closed-cycle refrigerator was used to control the temperature in the range of 300 K to 5 K. A 50-cm-long spectrograph equipped with a liquid-nitrogen-cooled charge-coupled device (CCD) was used. HeCd laser lines at 325 nm and 442 nm were used for the excitation of MAPbCl$_3$ and MAPbBr$_3$, respectively, and a 532 nm ND-YAG laser was used for MAPbI$_3$ excitation. For magnetic-field-dependent PL measurements, we used a capacitor-driven pulsed magnet located at the University of Tokyo and a cryogen-free 19 T superconducting dc magnet located at Tohoku University (Figure S1). The maximum field strength and transient time of the pulsed magnet were up to 60 T and $\sim$ 35 ms (Figure S2), respectively. For PL measurements in pulsed magnetic fields, an electron magnifying CCD (EMCCD) was employed to take PL spectra every 400 $\mu$s by using the spectra-kinetics acquisition mode during a field pulse. For polarization-dependent PL measurements, a thin plastic circular polarizer was inserted between the optical fiber and sample. By simply reversing the magnetic-field direction, the PL transition of the reverse spin orientation can be detected without changing the sample alignment. To maintain a constant sample temperature (4.2 K) from eddy-current heating during the transient magnetic pulse, the bottom part of the PL measurement probe (sample mount) was made by non-metallic components.

## III. RESULTS AND DISCUSSION

### A. Photoluminescence Spectra of MAPbX₃

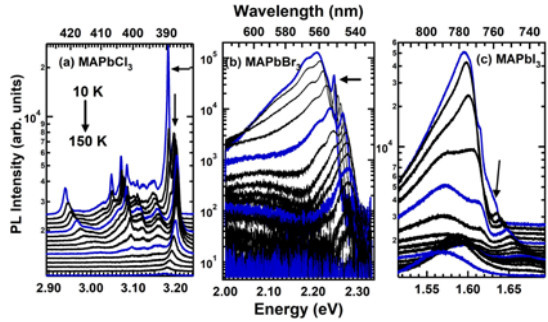

FIG. 1: Temperature dependence of the PL transition spectra of (a) MAPbCl₃, (b) MAPbBr₃, and (c) MAPbI₃ single crystals in steps of 10 K from 10 K to 150 K. Arrows indicate FX transitions.

Figure 1 displays the PL spectra of (a) MAPbCl₃, (b) MAPbBr₃, and (c) MAPbI₃ crystals at temperatures varying from 10 K to 150 K in steps of 10 K. All samples show band-edge FX transitions indicated by arrows and bound exciton (BX)transitions[25–27]. Both FX and BX transitions exhibit blue shifts as the temperature drops from 150 K to ∼ 50 K; with further drop in temperature, they exhibit red-shifts. The red-shift below 50 K is typical of excitons and caused by the increase of binding energy with decreasing temperature. The FX transition intensity changes sensitively with varying external excitation power[26].

Figure 2 shows PL spectra taken under different magnetic fields. Because the transitions of MAPbCl₃ in Figure 2a are in the near UV region, only unpolarized spectra can be obtained. At 0 T, MAPbCl₃ shows two FX transitions marked as peaks 1 and 2 that undergo spectral red-shifts in magnetic fields. Under high magnetic fields above 20 T, a new transition indicated by 3 emerges between the two FX transitions; this transition shows a blue-shift with further increase in the magnetic field.

Other sharp peaks in the lower-energy side of the FXs, which are believed to be impurity-related transitions, show red-shifts under magnetic fields. For the other two samples, MAPbBr₃ (b) and MAPbI₃ (c), we obtained circularly polarized PL spectra under magnetic fields, and both show blue-shifts. We confirmed that up to 50 T, the linearly polarized PL spectra for these samples did not show any appreciable difference between the longitudinal and transverse directions (see Figure S3 in Supplemental Information). In Figure 2b, MAPbBr₃ shows two energetically closely located FX transitions indicated by vertical arrows with peaks 1 and 2 in the inset of Figure 2b. Broad peaks numbered 3 and 4 in Figure 2b are BXs. Unlike MAPbCl₃ and MAPbBr₃, as shown in Figure 2c, MAPbI₃ exhibits a single FX transition (peak 1) with two BX transitions (peaks 2 and 3)[26]. In this case, the FX transition shows remarkable changes with increasing magnetic fields for both $\sigma^+$ and $\sigma^-$ directions.

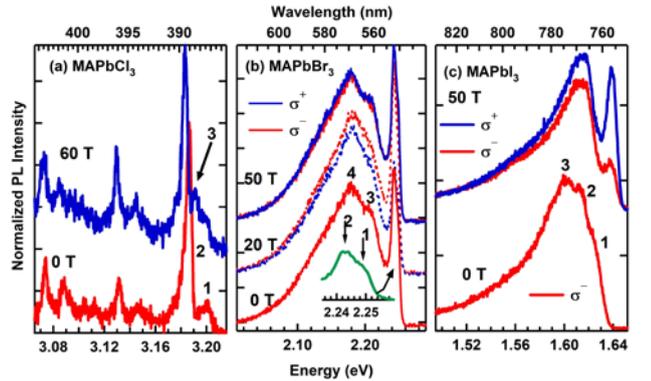

FIG. 2: PL spectra under magnetic fields. (a) For MAPbCl₃, two FX transitions labeled as peaks 1 and 2 show red-shifts, whereas peak 3 emerged above 20 T and exhibits a blue-shift. (b) For MAPbBr₃, two FX transitions, labeled peaks 1 and 2, with bound excitons (3 and 4) show blue-shifts between 0 to 50 T. (c) For MAPbI₃, the FX transition appeared as a small side peak labeled as peak 1 at B = 0 T, which notably changed in intensity under magnetic fields.

### B. Transition Energy of MAPbCl₃ in Magnetic Fields

Figure 3 shows the PL transition-energy shifts ($\Delta E$) of excitons under varying magnetic fields. The transition-energy shift $\Delta E$ of excitons under magnetic fields can be described as follows:

$$\Delta E = \pm \frac{1}{2}\Delta g \mu_B B + c_0 B^2 \quad (2)$$

The first term on the right side of Equation 2 is the Zeeman energy, wherein $\Delta g$ is the difference between the electron and hole $g$-factors (effective g-factor, here- after) and $\mu_B$ is the Bohr magneton. The second term is the diamagnetic shift with the diamagnetic coefficient $c_0 = e^2 a^{*2}_B/8\mu$, where $e$ is the electric charge, $a^*_B$ is the exciton Bohr radius, and $\mu$ is the exciton effective reduced mass. As shown in Figure 3, the fitting for peak 1 (red solid line) exhibits a negative diamagnetic coefficient of $c_0 = -0.50 \pm 0.034$ $\mu$eV/T² without the Zeeman term. Similarly, peak 3, which emerges above 20 T, has a positive sign with the same diamagnetic coefficient, $c_0 = +0.50 \pm 0.028$ $\mu$eV/T² (blue solid line). Peak 2 shows a different behavior from peaks 1 and 2. From 0 to ∼ 20 T, it decreases slowly, and then decrease rapidly with further increase in the magnetic fields. To analyze the PL energy shift of peak 2, we divide the fitting range into the ranges below and above 20 T. Between 0 and 20 T, the fitting follows $\Delta E = -1.08 \times 10^{-6} B^2$ (green solid line). Above 20 T, the fitting follows $\Delta E =$ $-(0.044)$ $\frac{1}{2}\mu_B B + 0.72 \times 10^{-6} B^2$ (black solid line). There- fore, the diamagnetic coefficient of peak 2 changes its sign from negative ($c_0 = -1.08 \pm 0.17$ $\mu$eV/T$^2$) to pos- itive ($c_0 = +0.72 \pm 0.06$ $\mu$eV/T$^2$) as the magnetic field increases beyond 20 T. This change in sign indicates that the effective reduced mass has also changed from nega- tive to positive. To understand this peculiar behavior of effective mass, one must consider an effective-mass cor- rection caused by the polaronic and SOI effects. The formation of polarons induced by the electron–phonon interaction decreases the electron energy by the polaron binding energy. In the effective-mass approximation, the reduction of the electron energy increases the effective mass. Consequently, the effective mass of the polaron is greater than that of the band electron. In addition to the polaron effect, the SOI effect modifies effective mass. The two-dimensional (2D) expression for the polaron ef- fective mass due to Rashba splitting at $k = 0$ can be expressed as follows[28]:

$$\frac{1}{m_{xx}^*} = \frac{1}{m^*} \pm \frac{\alpha k_y^2}{k^3},$$

where $m_{xx}^*$ is the polaron effective mass and $\alpha$ is the Rashba coupling constant. Although Equation 3 is es- timated for a 2D semiconductor, this equation can be applied to three-dimensional crystals in the presence of a magnetic field because the electron motion in the $z$- direction (parallel to the magnetic field) carries the lin- ear momentum, and the circular motion is limited in the $xy$ plane. Owing to the second term on the right side of Equation 3, the effective mass of the negative spin can be negative under the strong SOI. Therefore, the negative diamagnetic coefficients of peaks 1 and 2 (below 20 T) are associated with polarons in the negative-spin Rashba band.

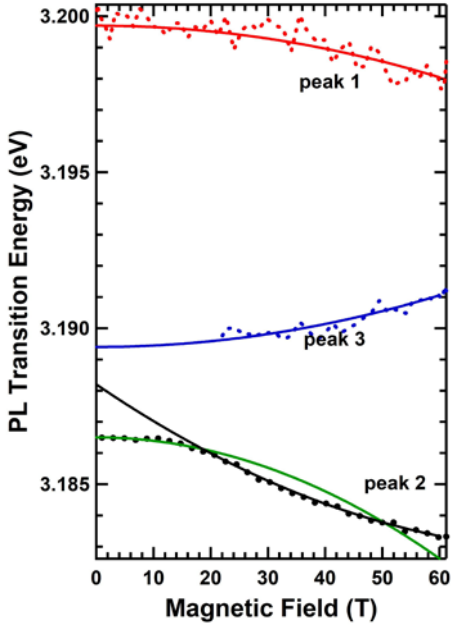

FIG. 3: PL transition-energy shift of three FX transitions from the MAPbCl$_3$ crystal (broken lines). Peaks 1 and 2 show red-shifts, whereas peak 3, which emerged above 20 T, exhibits a blue-shift. Peaks 1 and 3 fit with $\pm c_0 B^2$, and the fits are shown as red and blue solid lines, respectively. Peak 2 follows $-B^2$ fitting below 20 T; above 20 T, it follows the fitting equation $-\frac{1}{2}\Delta g\mu_B B + c_0 B^2$.

When the cyclotron frequency becomes higher than the polaron frequency, carrier motion decouples from the polar lattice[8]. The cyclotron ($\omega_c$) and polaron ($\omega_p$) frequencies[29] are given by

$$r_c = \frac{\left(\hbar^2/2m^*\right)^{1/2}}{\hbar^2\omega_c} = \frac{f_B}{\sqrt{2}}, \tag{4}$$

$$r_p = \frac{\left(\hbar^2/2m^*\right)^{1/2}}{\hbar^2\omega_p}, \tag{5}$$

$$\frac{\omega_c}{\omega_p} = \left(\frac{r_p}{r_c}\right)^2 = 2\left(\frac{r_p}{f_B}\right)^2, \tag{6}$$

where $r_c$ ($r_p$) is the cyclotron (polaron) radius and $f_B = \sqrt{\hbar/eB} = \sqrt{2}r_c$ is the magnetic length. A comparison of the cyclotron and polaron frequencies indicates that carrier motion decouples from the polar lattice when the magnetic length $f_B$ is comparable to the polaron radius. At 20 T, the cyclotron radius $r_c = 4.0$ nm ($f_B = 5.7$ nm), which is close to the radius of the large polaron in MAPbCl$_3$[30]. Therefore, at approximately 20 T, the po- laron effect can be softened. For the SOI-induced Rashba effect, under low magnetic fields, the Rashba effect dom- inates the Zeeman effect and vice versa under high mag- netic fields[31–33]. As a consequence of the softened po- laron and Rashba effects, the slope of the peak-2 transi- tion changes, and the effective mass changes from neg- ative to positive as the magnetic field increases beyond 20 T. The energy-shift slope of peak 1 does not change, and the negative effective mass is maintained in the en- tire field range up to 60 T. The magnetic length at 60 T is $f_B = 3.3$ nm, which is not small enough to decouple the electron from the polar lattice in the case of a small polaron that is confined within a unit cell with the lat- tice constant $a = 5.67$ Å for MAPbCl$_3$. Therefore, the peak-1 and peak-3 transitions are associated with small polarons in the negative and positive spin Rashba bands, respectively. Although the self-trapped small polaron is difficult to form in 3D OIHP materials, our MPL tran- sition suggests that the formation of the small polaron is possible in the MAPbCl$_3$ crystal because of the large ionic bonding ratio and small lattice constant.

## C. Transition Energy of MAPbBr$_3$ in Magnetic Fields

For the MAPbBr$_3$ peak transition energies in Figure 4, broken and solid lines correspond to the PL transition-energy shifts of peaks 1 and 2 from Figure 2b inset, respectively, and blue and red indicate $\sigma^+$ and $\sigma^-$ from Figure 2b, respectively. Because the electronegativity of Br is smaller than that of Cl, the polaronic effect of MAPbBr$_3$ is weaker than that of MAPbCl$_3$. Consequently, the PL transition behavior of MAPbBr$_3$ is expected to be different from that of MAPbCl$_3$. The total transition-energy shifts in the magnetic fields of two FX transitions are almost identical in the same polarization directions. $\sigma^-$ transitions (red solid and broken lines) follow Equation 2 with $\Delta g = -1.82$ and $c_0 = +1.54 \pm 0.045$ $\mu$eV/T$^2$. These values slightly deviate from the result previously published by Tanaka et al.[20] who reported these values as $\Delta g = \pm 2.03$ and $c_0 = +1.28$ $\mu$eV/T$^2$ from magneto-absorption experiments. However, in the opposite spin direction ($\sigma^+$), the PL transition-energy shifts for peaks 1 and 2 do not move in magnetic fields below $\sim 30$ T. Above 30 T, the shifts fit Equation 2 with $\Delta g = +0.78$ and $c_0 = +0.46 \pm 0.07$ $\mu$eV/T$^2$, which are significantly smaller than those for the opposite ($\sigma^-$) direction. The difference between the different polarization directions may be due to the effective mass corrections indicated in Eq 3. Under low magnetic fields, both the Rashba and polaron effects are dominant. However, with increasing magnetic fields, the Zeeman energy becomes more important than the Rashba effect, and the cyclotron frequency is comparable to the polaron frequency. Therefore, under high magnetic fields, both the Rashba and polaron effects are suppressed and the cyclotron motion becomes predominant above 30 T.

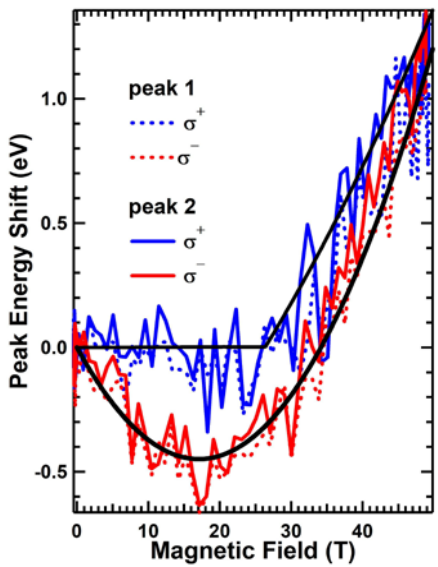

FIG. 4: Two FX transition-energy shifts of the MAPbBr$_3$ crystal in different spin directions. Blue broken (peak 1) and solid (peak 2) lines are from $\sigma^+$ transitions, and red broken (peak 1) and solid (peak 2) lines are associated with $\sigma^-$ transitions. The energy shifts of peaks 1 and 2 in the same direction are almost equal. Black solid lines are fitting lines. A straight line for $\sigma^+$ below 28 T is a guide for the eye.

### D. Transition Energy of MAPbCl$_x$Br$_{3-x}$ in Magnetic Fields

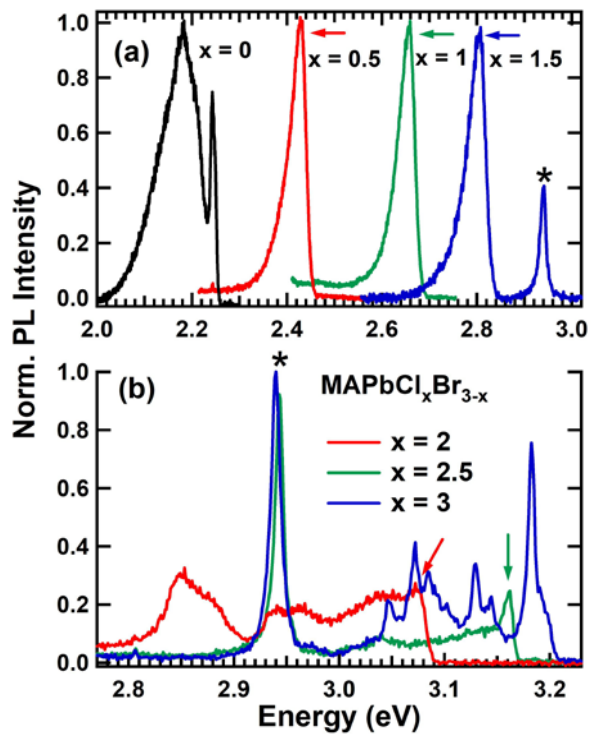

FIG. 5: PL spectra of MAPbCl$_x$Br$_{3-x}$ at $B = 0$ T, 4.2 K. (a) $x = 0$, 0.5, 1, 1.5 and (b) $x = 2$, 2.5, 3.

Because the difference of the PL spectral shapes between MAPbCl$_3$ and MAPbBr$_3$ is large (see Figure 1 a and b), we attempted to measure unpolarized PL measurements for MAPbCl$_x$Br$_{3-x}$ ($x = 0.5 \sim 2.5$) crystals in magnetic fields. As seen in Figure 5a, for $x = 0.5 \sim 1.5$, PL spectral shapes are similar to MAPbBr$_3$, whereas for $x = 2$ and 2.5 in Figure 5b, spectral shapes are similar to MAPbCl$_3$. When $x \geq 1.5$, there is a common peak that appeared at $\sim 2.93$ eV marked by asterisks, which can be a bound exciton associated with Cl.

In the presence of magnetic field in Figure 6, $x = 0.5$ and (red markers) and 1.0 (green markers) exhibit typical diamagnetic energy shift, $\Delta E = c_0 B^2$ with the diamagnetic coefficients $c_0 = 0.690 \pm 0.010 \mu$eV/T$^2$ and $0.370 \pm 0.007$ $\mu$eV/T$^2$ for $x = 0.5$ and 1.0, respectively. For the case of $x = 1.5$, the peak energy does not move (see zero line). With further increasing $x$, the diamagnetic energy shift becomes negative. Below 20 T, both peak transition energies fit well with the diamagnetic co-efficients $c_0 = -4.959 \pm 0.30 \mu$eV/T$^2$ and $-1.741 \pm 0.19$ $\mu$eV/T$^2$ for $x = 2.0$ and 2.5, respectively (see black solid lines). Comparing not only for spectral shapes but also energy shifts in magnetic fields, when $x$ is smaller or greater than 1.5, the PL transition characteristics become MAPbBr$_3$-like or MAPbCl$_3$-like behavior, respectively.

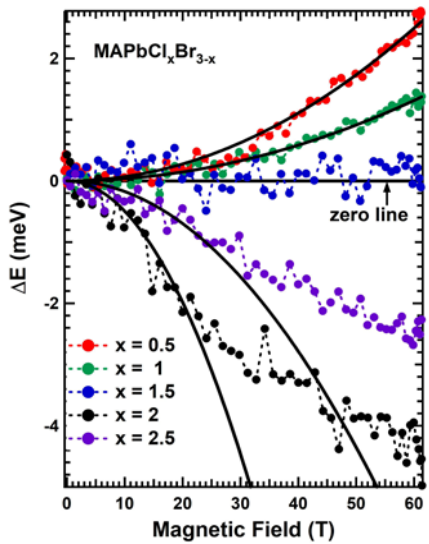

FIG. 6: Estimated peak transition energy in magnetic fields for $x = 0.5 \sim 2.5$ indicated by arrows in Fig. 5 . Two samples, $x = 0.5$ (red) and 1.0 (green), follows the diamagnetic fitting equation $\Delta E = c_0 B^2$ (black solid lines). For $x = 1.5$ (blue), the peak transition energy barely moves. When $x$ exceed 2, the diamagnetic energy shift becomes negative (black and purple).

### E. Transition Energy of MAPbI$_3$ in Magnetic Fields

Iodine has the smallest electronegativity among the halogens used in this study. Therefore, its polaronic ef- fect might be limited in comparison with other elements. The PL transition-energy shifts of MAPbI$_3$ exhibit a completely different behavior from those of the other samples, as shown in Figure 7. Neither of the $\sigma^{\pm}$ transi- tions fit Equation 2, and both exhibit $\Delta E = aB + bB^m$ dependency. In the figure, green lines indicate the fit- ted values; for $\sigma^+$, $a(\sigma^+) = 0.1270 \pm 0.0128$, $b(\sigma^+) = 0.79537 \pm 0.0341$, and $m(\sigma^+) = 0.50328 \pm 0.0373$, and for $\sigma^-$, $a(\sigma^-) = 0.17591 \pm 0.0157$, $b(\sigma^-) = 0.34397 \pm 0.0437$, and $m(\sigma^-) = 0.49854 \pm 0.109$. Considering the power $m$, which is close to 0.5, both polarizations follow $B$ under low magnetic fields below 20 T and are linear in $B$ under high magnetic fields above 20 T. Furthermore, the room-temperature MPL transition behavior shows the same $B$ behavior below 20 T. Such a non-linear energy transition under low magnetic fields may be due to the band non-parabolicity combined with the polaron effect[34–36]. Peeters et al.[35] reported that the non-linear transition occurs at the anticrossing region of the $(n+1)$th zero-phonon and $n$th Landau level. The LO phonon energy in MAPbI$_3$ is known to be $\sim 35$ eV; in our case, the total transition energy between 0 to 50 T is approximately 12 meV. Therefore, the anti-crossing be-tween the zero-phonon and no-phonon scenarios does not meet our case.

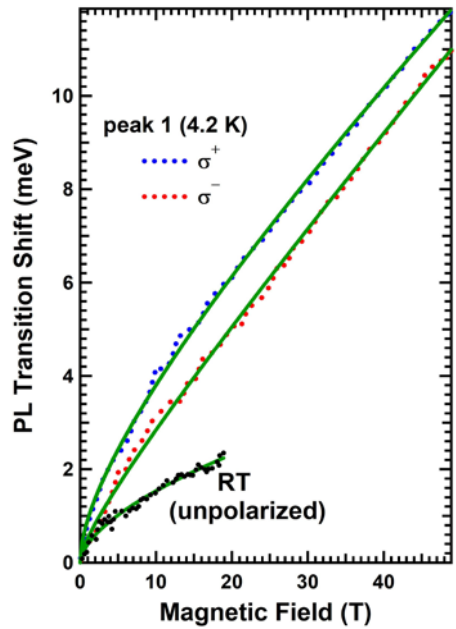

FIG. 7: FX transition-energy shift of the MAPbI$_3$ crystal in different spin directions. Blue and red and broken lines are associated with $\sigma^+$ and $\sigma^-$ transitions, respectively. Both peaks show a power dependency ($\Delta E = aB + bB^m$, where a and b are fitting parameters). A transition at room temperature (black dotted lines), measured under a magnetic field of 19 T produced by a superconducting magnet, also follows the same fitting equation (black solid line).

As chlorine is substituted with iodine, the energy gap decreases, and thus the bandgap non-parabolicity be-comes important. For a non-parabolic band, the polaron effective reduced mass ($\mu$) is no longer a constant value; rather, it is a function of the magnetic field[37]:

$$\Delta \mu^*(B) = \frac{eB}{\Delta E},  \tag{7}$$

where $\Delta \mu^*(B)$ and $\Delta E$ are the changes of effective re-duced mass and the energy shift in magnetic fields, re-spectively. Figure 8 shows the change in effective reduced mass under magnetic fields based on Equation 7 in units of the electron rest mass ($m_0$), where broken lines indi- cate experimental values and green solid lines indicate fit- ted values. For the low-temperature transitions at 4.2 K, the reduced effective mass increases rapidly below 20 T as the MPL transition energies increase rapidly, as shown in Figure 5. In the high field region above 20 T, the change in slope of the effective reduced mass becomes low and shows saturation behavior, and the transition energies are linear in B . At room temperature, the effec- tive reduced mass is larger than that at low temperatures (black dotted markers). This temperature dependence of the change in effective reduced mass may be due to the difference in lattice structures between room temperature and 4.2 K.

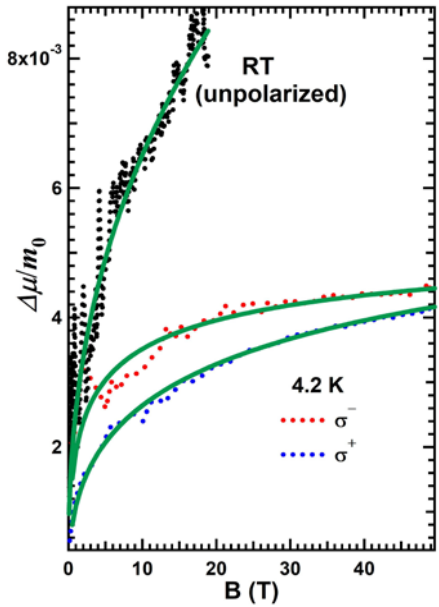

FIG. 8: Estimated change in effective reduced mass ($\Delta\mu^*(B)$) under magnetic fields in units of the electron rest mass ($m_0$). The non-linear MPL transition is due to the change in the magnetic-field-dependent effective reduced mass.

## IV. CONCLUSION

In conclusion, we obtained MPL spectra from MAPbX$_3$ (X = Cl, Br, and I) under strong magnetic fields of up to 60 T by using pulsed magnets. Because the halide changes the MPL transitions in an inconsistent manner, we could not find a unified model to explain the MPL transition-energy behaviors. For MAPbCl$_3$ crys- tal, two MPL transition peaks follow $B^2$ behavior up to 60 T at 4.2 K. These transitions are attributed to small polaron transitions. One transition peak energy shows a quadratic dependency on $B$ with a negative sign below 20 T and then follows a normal diamagnetic shift above 20 T. This behavior can be understood as a combination of the Rashba and the large polaron ef- fects, which cause a negative effective reduced mass at low fields below 20 T. The large polaron decouples when the cyclotron frequency becomes comparable to the po- laron frequency with increasing magnetic fields above 20 T. For MAPbBr$_3$ crystal, the MPL transition-energy be- havior follows a normal diamagnetic shift. However, the coefficients are different from previously reported val- ues obtained using magneto-absorption measurements. This is because the MPL transition may reflect the po- laron effect, whereas the magneto-optical absorption is solely associated with the band-edge transition. For MAPbCl$_x$Br$_{3-x}$ samples, when $x \leq 1.5$ or $x \geq 2.0$, the PL transition characteristics are similar to MAPbBr$_x$ or MAPbCl$_3$, respectively.

The power-law dependency of the MPL transition en- ergies for MAPbI$_3$ under magnetic fields associates with the band non-parabolicity. Because of the non-parabolic band effect combined with the polaron effect, the effec- tive reduced mass changes with the magnetic field. Con- sequently, by replacing the halogen in OIHP crystals, it is not possible to explain the MPL transition behavior in a unified model, and one must consider different effects with different halogens in OIHP materials.

## V. ACKNOWLEDGEMENTS

The work at Dankook University is supported by the National Research Foundation of Korea (NRF-2016R1D1A1B01006437) funded by the Korea government (MSIT). The research conducted at SKKU is sup- ported by the National Research Foundation of Korea (NRF-2019M3D1A1078304) funded by the Ministry of Science and ICT. A part of this work has been supported by ICC-IMR and GIMRT program of IMR, Tohoku Univ. and the visiting professor program of ISSP, The Univ. of Tokyo.

[1] National Renewable Energy Labora- tory Best Research-Cell Efficiency Chart. https://www.nrel.gov/pv/assets/pdfs/best-research- cell-efficiencies.20200104.pdf (accessed on February 2021).

[2] M. Jeong, I. W. Choi, E. M. Go, Y. Cho, M. Kim, B. Lee, S. Jeong, Y. Jo, H. W. Choi, J. Lee, J.-H. Bae, S. K. Kwak, D. S. Kim, and C. Yang, Science 2020, 369, 1615.

[3] Z. G. Yu, Sci. Rep. 6, 28576 (2016).

[4] D. Canneson, E. V. Shornikova, D. R. Yakovlev, T. Rogge, A. A. Mitioglu, M. V. Ballottin, P. C. M. Chris- tianen, E. Lhuillier, M. Bayer, and L. Biadala, Nano Lett. 17, 6177 (2017).

[5] M. A. Becker, R. Vaxenburg, G. Nedelcu, P. C. Sercel, A. Shabaev, M. J. Mehl, J. G. Michopoulos, S. G. Lam- brakos, N. Bernstein, J. L. Lyons, T. Stöferle1, R. F. Mahrt, M. V. Kovalenko, D. J. Norris, G. Rainò and A. L. Efros, Nature 553, 189 (2018).

[6] M. Isarov, L. Z. Tan, M. I. Bodnarchuk, M. V. Ko- valenko, A. M. Rappe, and E. Lifshitz, Nano Lett. 17, 5020 (2017).

[7] H. Ryu, D. Y. Park, K. M. McCall, H. R. Byun, Y. Lee, T. J. Kim, M. S. Jeong, J. Kim, M. G. Kanatzidis, and J. I. Jang, J. Am. Chem. Soc. 142 (50), 21059–21067 (2020).

[8] M. Baranowski and P. Plochocka, Adv. Energy Mater. 10, 1903659 (2020).

[9] D. Meggiolaro, F. Ambrosio, E. Mosconi, A. Mahata, and F. De Angelis, Adv. Energy Mater. 10, 1902748 (2020).

[10] D. Ghosh, E. Welch, A. J. Neukirch, A. Zakhidov, and S. Tretiak, J. Phys. Chem. Lett. **11**, 3271 (2020).

[11] A. R. S. Kandada and C. Silva, J. Phys. Chem. Lett. **11**, 3173 (2020).

[12] A. J. Neukirch, W. Nie, J. C. Blancon, K. Appavoo, H. Tsai, M. Y. Sfeir, C. Katan, L. Pedesseau, J. Even, J. J. Crochet, G. Gupta, A. D. Mohite, and S. Tretiak, Nano Lett. **16**, 3809 (2016).

[13] M. Park, A. J. Neukirch, S. E. Reyes-Lillo, M. Lai, S. R. Ellis, D. Dietze, J. B. Neaton, P. Yang, S. Tretiak, and R. A. Mathies, Nat. Commun. **9**, 2525 (2018).

[14] E. Menéndez-Proupin, C. L. B. Beltrán Ríos, and P. Wahnón, Phys. Status Solidi RRL **9**, 559 (2015).

[15] X.-Y. Zhu and V. Podzorov, J. Phys. Chem. Lett. **6**, 4758 (2015).

[16] A. M. Soufiani, F. Huang, P. Reece, R. Sheng, A. Ho-Baillie, and M. A. Green, Appl. Phys. Lett. **107**, 231902 (2015).

[17] K. Miyata, D. Meggiolaro, M. T. Trinh, P. P. Joshi, E. Mosconi, S. C. Jones, F. De Angelis, and X.-Y. Zhu, Sci. Adv. **3**, e1701217 (2017).

[18] R. G. Niemann, A. G. Kontos, D. Palles, E. I. Kamitsos, A. Kaltzoglou, F. Brivio, P. Falaras, and P. J. Cameron, J. Phys. Chem. C **120**, 2509 (2016).

[19] T. Glaser, C. Müller, M. Sendner, C. Krekeler, O. E. Semonin, T. D. Hull, O. Yaffe, J. S. Owen, W. Kowalsky, A. Pucci, and R. Lovrinčić, J. Phys. Chem. Lett. **6**, 2913 (2015).

[20] K. Tanaka, T. Takahashi, T. Ban, T. Kondo, K. Uchida, and N. Miura, Solid State Commun. **127**, 619 (2003).

[21] K. Galkowski, A. Mitioglu, A. Miyata, P. Plochocka, O. Portugall, G. E. Eperon, J. T.-W. Wang, T. Stergiopoulos, S. D. Stranks, H. J. Snaith, and R. J. Nicholas, Energy Environ. Sci. **9**, 962 (2016).

[22] A. Miyata, A. Mitioglu, P. Plochocka, O. Portugall, J. T.-W. Wang, S. D. Stranks, H. J. Snaith, and R. J. Nicholas, Nat. Phys. **11**, 582 (2015).

[23] G. Maculan, A. D. Sheikh, A. L. Abdelhady, M. I. Saidaminov, Md A. Haque, B. Murali, E. Alarousu, O. F. Mohammed, T. Wu, and O. M. Bakr, J. Phys. Chem. Lett. **6**, 3781 (2015).

[24] M. I. Saidaminov, A. L. Abdelhady, B. Murali, E. Alarousu, V. M. Burlakov, W. Peng, I. Dursun, L. Wang, Y. He, G. Maculan, A. Goriely, T. Wu, O. F. Mohammed, and O. M. Bakr, Nat. Commun. **6**, 7586 (2015).

[25] L. Q. Phuong, Y. Nakaike, A. Wakamiya, and Y. Kanemitsu, J. Phys. Chem. Lett. **7**, 4905 (2016).

[26] H.-J. Jo, D. Y. Park, M. G. So, Y. Kim, J. S. Kim, and M. S. Jeong, Current Appl. Phys. **19**, 60 (2019).

[27] J. Tilchin, D. N. Dirin, G. I. Maikov, A. Sashchiuk, M. V. Kovalenko, and E. Lifshitz, ACS Nano **10**, 6363 (2016).

[28] Z. Li, Z. Ma, A. R. Wright, and C. Zhang, Appl. Phys. Lett. **90**, 112103 (2007).

[29] N. Miura, Physics of Semiconductors in High Magnetic Fields, Oxford Science Publications, Oxford University Press, Oxford, GB (2008).

[30] M. Sendner, P. K. Nayak, D. A. Egger, S. Beck, C. Müller, B. Epding, W. Kowalsky, L. Kronik, H. J. Snaith, A. Pucci, and R. Lovrinčić, Mater. Horiz. **3**, 613 (2016).

[31] M. H. Weiler, Semiconductors and Semimetals **16**, 119 (1981).

[32] T. Y. Kim, S. Joo, J. Lee, J. H. Suh, S. Cho, S. U. Kim, K. Rhie, J. Hong, and K.-H. Shin, J. Kor. Phys. **54**, 697 (2009).

[33] P. Pfeffer and W. Zawadzki, Phys. Rev. B **68**, 035315 (2003).

[34] D. M. Larsen, Phys. Rev. B **30**, 4595 (1984).

[35] F. M. Peeters and J. T. Devreese, Phys. Rev. B **31**, 3689 (1985).

[36] J. Singleton, R. J. Nicholas, D. C. Rogers, and C. T. B. Foxon, Surf. Sci. **196**, 429 (1988).

[37] E. D. Palik, G. S. Picus, S. Teitler, and R. F. Wallis, Phys. Rev. **122**, 475 (1961).

Y. H. Shin,[1] Halim Choi,[1] C. Park,[1] Yongmin Kim,[1*] D. Park,[2] M. S. Jeong,[2] H. Nojiri,[3] Z. Yang,[4] and Y. Kohama[4]

[1]*Department of Physics, Dankook University, Cheonan 31116, Korea*
[2]*Department of Physics, Hanyang university, Seoul 04763, Korea*
[3]*Institute for Materials Research, Tohoku University, Sendai 980-8577, Japan and*
[4]*Institute for Solid State Physics, The University of Tokyo, Kashiwa 277-8581, Japan*

(Dated: April 29, 2021)

## SUPPLEMENTAL INFORMATION

For MPL measurements, a standard single optical fiber technique was employed (Fig. S1). The CCD acquisition was triggered 5 ms prior to the magnetic field pulse. As seen in Fig. S2, the field rising and falling time is 13 ms and 22 ms, respectively, for 60 T magnet, and we an- alyzed MPL spectra that were taken during the falling magnetic fields in order to minimize the magnetic field variation during the spectral acquisition. For linearly po- larized MPL measurements, we obtained identical tran- sition behavior in both $\pi^{\pm}$ directions (Fig. S3). Circular polarization degrees for MAPbBr$_3$ and MAPbI$_3$ were es- timated over 40 % at 50 T (Figs. S5 and S6).

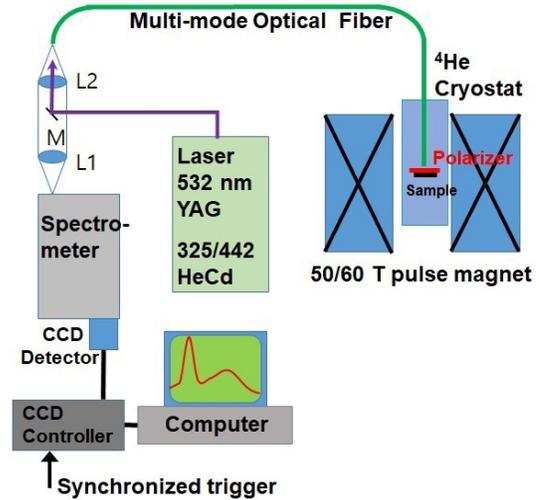

Figure S1. Experimental setup for photoluminescence under pulsed magnetic fields.

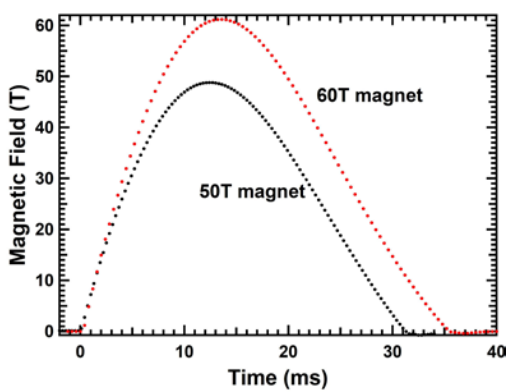

Figure S2. Pulsed magnetic field profiles used for this study. The total transient time of the fields are 31 ms and 35 ms for 50 T and 60 T magnets, respectively. Photoluminescence spectra were taken at each points, which is 400 μs acquisition time by using an EMCCD.

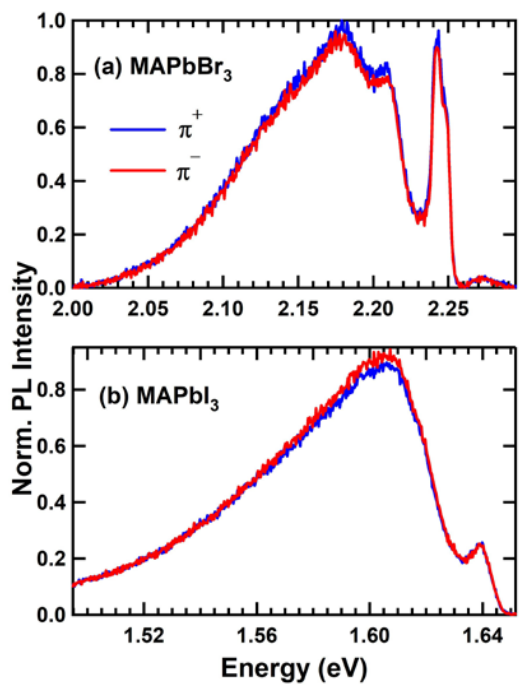

Figure S3. (a) Linearly polarised PL spectra at 50 T for MAPbBr₃ (b) and MAPbI₃. Both samples did not show any appreciable differences between $\pi^+$ and $\pi^-$ directions. The same field independent linear polarization effect was observed in the entire magnetic field range.

.

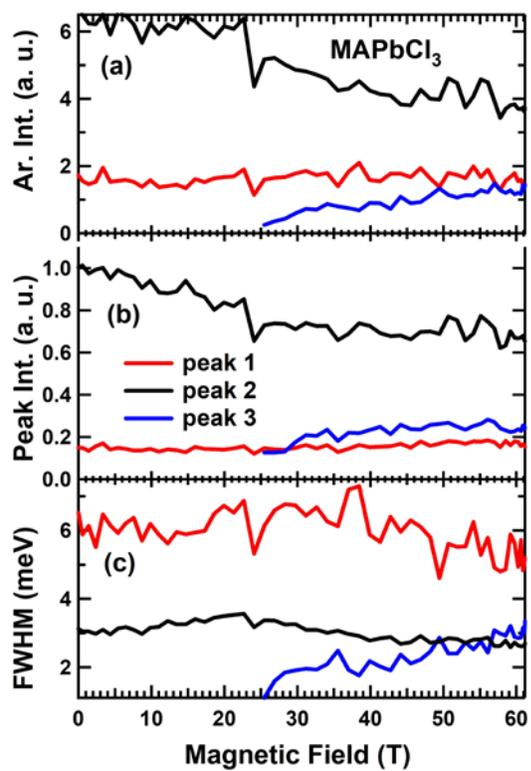

Figure S4. (a) Integrated areal PL intensity, (b) peak PL intensity and (c) the full width and half maximum (FWHM) in magnetic fields for MAPbCl$_3$.

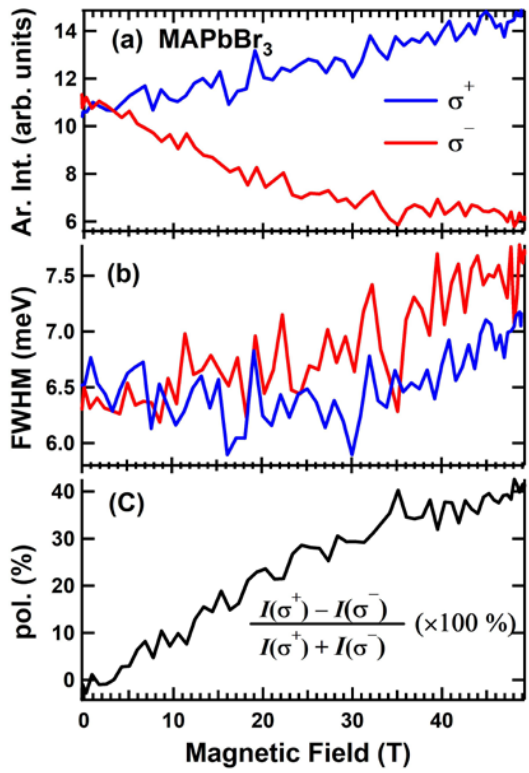

Figure S5. (a) Integrated areal PL intensity, (b) the full width and half maximum (FWHM) and (c) the total polarization in magnetic fields for MAPbBr₃.

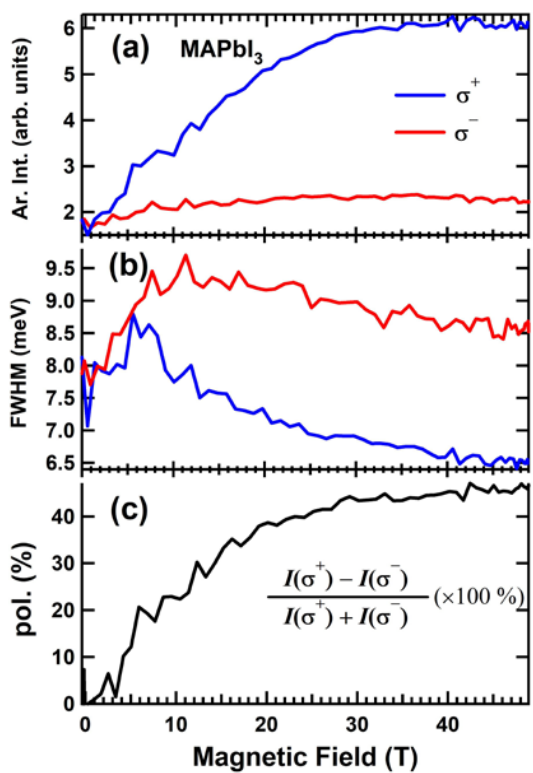

Figure S6. (a) Integrated areal PL intensity, (b) the full width and half maximum (FWHM) and (c) the total polarization in magnetic fields for MAPbI₃.